\documentclass[oldversion]{aa}  

\usepackage{graphicx}
\usepackage{txfonts}
\usepackage{float}

\def\N_EFEDS{2060}

\begin{document}

   \title{The eROSITA Final Equatorial-Depth Survey (eFEDS)}

   \subtitle{The stellar counterparts of eROSITA sources identified by machine learning and Bayesian algorithms}

   \author{P. C. Schneider\inst{1}
          \and
          S. Freund\inst{1}
          \and
          S. Czesla\inst{1}
          \and
          J. Robrade\inst{1}
          \and
          M. Salvato\inst{2}
          \and J. H. M. M. Schmitt\inst{1}
          }

   \institute{Hamburger Sternwarte, Gojenbergsweg 112, D-21029 Hamburg, Germany \email{astro@pcschneider.eu}
   \and
   Max-Planck-Institut für extraterrestrische Physik, Giessenbachstrasse 1, 85748 Garching, Germany
             }

   \date{Received ?; accepted ?}

 
  \abstract{Stars are ubiquitous X-ray emitters and will be a substantial 
  fraction of the X-ray sources detected in the on-going  all-sky survey performed 
  by the eROSITA instrument 
  aboard the Spectrum Roentgen Gamma (SRG) observatory. We use the X-ray sources in the 
  eROSITA Final Equatorial-Depth Survey (eFEDS) field observed during the SRG  performance 
  verification phase to investigate different strategies to identify the stars
  among other source categories. We focus here on 
  Support Vector Machine (SVM) and Bayesian approaches, and our approaches 
  are based on a cross-match  with the Gaia catalog, which will eventually 
  contain counterparts to virtually all stellar eROSITA sources. 
  We estimate that 2060 stars are among the eFEDS sources based on the 
  geometric match distance distribution, and we identify the 
  2060 most likely stellar sources with the SVM and Bayesian methods, 
  the latter being named HamStars in the eROSITA context. 
  Both methods reach completeness and reliability percentages of  
  almost 90\,\%, and the agreement between both methods is, incidentally, 
  also about 90\,\%. Knowing the true number of stellar sources allowed us
  to derive association probabilities $p_{ij}$ for the SVM method similar to the 
  Bayesian method so that one can construct samples with defined 
  completeness and reliability properties using appropriate cuts in 
  $p_{ij}$. The thus identified stellar sources show the typical characteristics
  known for magnetically active stars, specifically, they are generally compatible
  with the saturation level, show a large spread in activity for stars of spectral 
  F to G, and have comparatively high fractional X-ray luminosities for later spectral types.}


   \maketitle
%

\section{Introduction \label{sect:intro}\label{sect:LxLbol}}
Stars with convective envelopes, that is, with stellar 
masses between 0.08 and 1.85\,$M_\odot$ (spectral types M to mid A) 
show magnetic activity and
possess a corona emitting soft X-rays ($\lambda\sim1-100$\,\AA). 
Stars of the same mass, however, can have very different X-ray
properties, primarily depending on the stellar rotation period.
The largest ratios between X-ray ($L_X$) and bolometric luminosities ($L_{bol}$)
are observed for rapidly rotating young stars, which show $L_X/L_{bol}$ ratios 
close to the so-called saturation limit of $\log L_X/L_{bol}\sim-3$ \citep{Vilhu_1984,Pizzolato_2003}. On the other hand, old slowly rotating stars may
have $\log L_X/L_{bol}\sim-8$ \citep[see reviews by][]{Guedel_2004,Testa_2015}. 
Since stars spin down with age, the stellar X-ray luminosity also declines 
with age \citep{Skumanich_1972}. Therefore, X-ray surveys are most sensitive 
to young stars and contain comparatively few old stars, hence, the parameter
space of stellar activity is very unevenly sampled. 

In addition to this inherent bias toward active stars, existing stellar 
samples with well characterized  X-ray properties are relatively small,
typically  $\lesssim1\,000$ objects \citep{Schmitt_2004,Wright_2011, Freund_2018},
compared to other stellar samples with hundreds of thousands 
of stars such as RAVE \citep{Steinmetz_2006}, the Gaia-ESO 
survey \citep{Gilmore_2012}, or even billions of stars \citep[][]{Gaia_2016}. The eROSITA all-sky survey, 
described in \citet[][]{Predehl_2021} and designated as eRASS:8, is expected to provide 
detections of almost $10^6$\,stars \citep[see][]{Merloni_2012}, thus bringing
the sample count of X-ray emitting stars on par with other samples.

To harvest the full potential for stellar science, one needs to identify
the stars among the other X-ray emitting objects in the eROSITA source list,
that is, a classification task. The final data of the eROSITA all-sky survey
will not be available before 2024, however, the eROSITA Final Equatorial 
Depth Survey (eFEDS) already provides the 
X-ray sensitivity expected after the completion of the all-sky survey 
for a field of $\sim 140$ square degrees  
\citep[][submitted to A\&A]{Brunner_2021, Salvato_2021}. Hence, eFEDS
provides an excellent opportunity to 
develop the methods required for the identification task.

The classification of large numbers of objects into different categories 
based on their measured properties is an old task and
has been approached by different mathematical methods such as frequentist 
\citep[e.g.,][]{Fischer_1938} or Bayesian procedures \citep{Binder_1978}. 
Nowadays, machine learning (ML) approaches are also popular thanks to improving algorithms, 
increasing computing power, and large datasets. These techniques are now regularly applied in 
the astrophysical context (e.g., \citet{Marton_2019, Vioque_2020}, and \citet{Melton_2020} used ML techniques 
to identify young stars).
Here, we present a Support Vector Machine (SVM) and a Bayesian method to
identify stellar X-ray sources within the eROSITA source catalog.

The paper is 
structured as follows. We translate the task of identifying the stars 
into a classification problem in sect.~\ref{sect:identification}, present
the SVM and Bayesian approaches in sects.~\ref{sect:SVM} and \ref{sect:Bayesian},
compare their results in sect.~\ref{sect:comparison}, and provide our conclusions
as well as the outlook in sect.~\ref{sect:conclusions}.

\section{The identification task \label{sect:identification}}
To identify and characterize the stellar eROSITA sources,
we need information from other wavelengths in addition to the X-ray 
data. Specifically, the identification of the stellar content in eROSITA
is based on matching the eROSITA sources to 
a catalog containing only eligible stellar counterparts, i.e., stars that 
may emit X-rays at flux levels detectable by eROSITA. An eROSITA source $i$
is classified as stellar if an association between eROSITA source $i$ and
a counterpart $j$ exists that  
has properties compatible with the hypothesis 
that counterpart $j$ is responsible for the X-ray emission, that is,
if one or more reasonable stellar counterparts ($j_1, \dots, j_n$)
exist so that the association $i\leftrightarrow j_1$ (or $i\leftrightarrow j_2$, etc.)
is likely while taking the possibility of chance alignments into account, too.

This approach differs from many other catalog cross-matching approaches in 
the sense that a complete identification of the eROSITA detections 
is not required: stars and non-stars are the only two relevant source 
categories for us, and only for objects in the first category do we attempt to find 
the correct counterpart (the problem of the identification, characterization and 
classification of the full eFEDS sample is discussed in Salvato et al., subm.). 
The non-stars category 
includes other source categories,
mainly AGN (Liu et al., submitted) and nearby galaxies (Vulic et al., submitted),
but also spurious X-ray detections, for example, due to background fluctuations.
Objects in the nonstellar category are here treated as 
random associations and considered in a statistical sense. 
In summary, the categorization of the eROSITA sources is deferred to 
classifying the associations $i \leftrightarrow j$ between eROSITA sources ($i$)
and eligible stellar counterparts ($j$) as probable with respect to the alternative
that the association is spurious, i.e., that object $j$ in the match catalog is not
responsible for X-ray source $i$.

\begin{figure}
  \centering
  \includegraphics[width=0.49\textwidth]{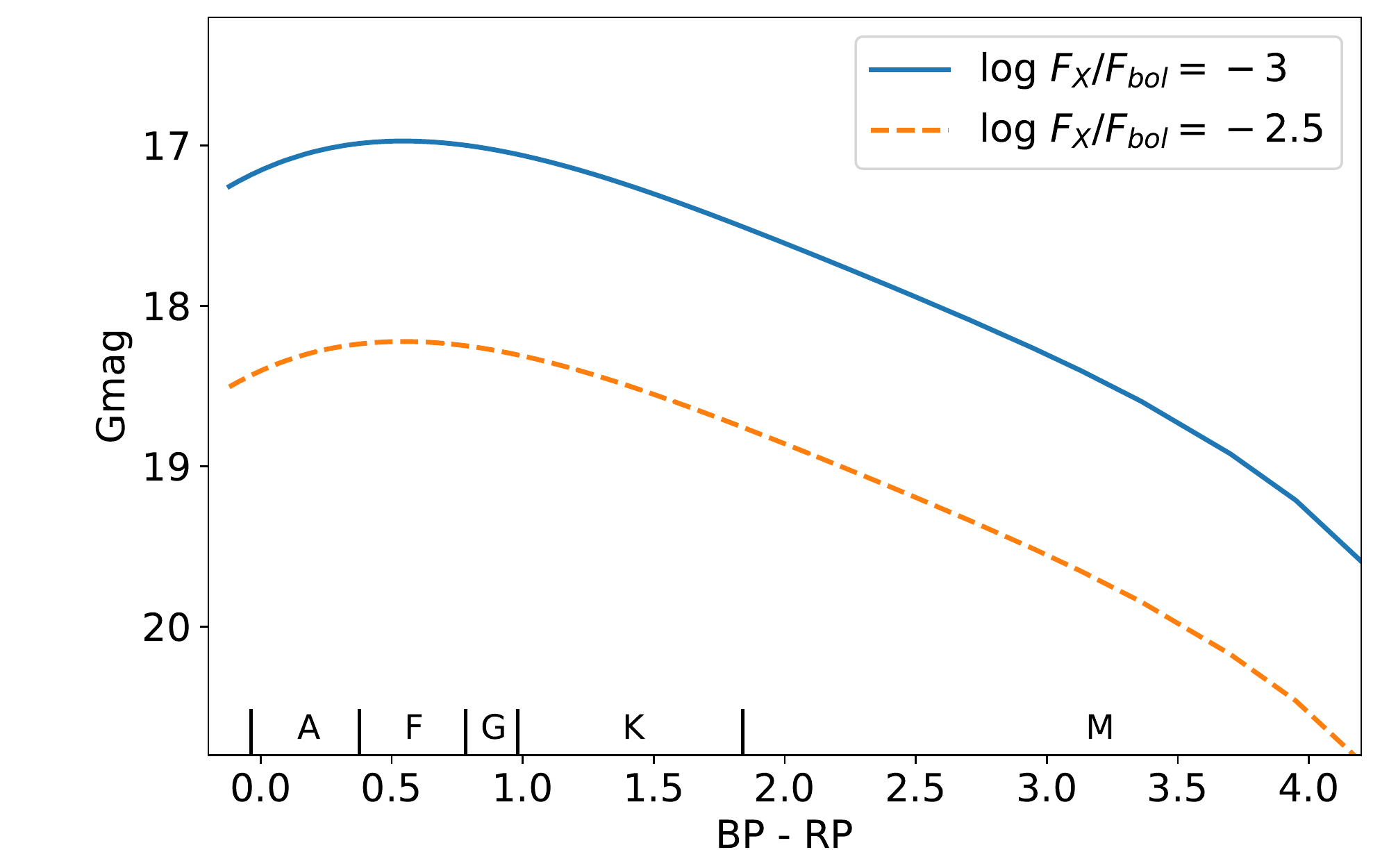}
  \caption{Limiting G~magnitude for stars emitting at X-ray activity levels of $\log L_X/L_{bol}=-2.5$ and -3 
  assuming the average eFEDS exposure depth.
  \label{fig:limG}}
\end{figure}

In this scheme, the match catalog containing the eligible stellar counterparts 
is crucially important for the classification and must include counterparts to
virtually all stellar eROSITA sources but may lack any other source
category. The saturation 
limit for stars of $\log L_X/L_{bol}\approx-3$ \citep{Pizzolato_2003, Wright_2011} implies a 
limiting optical magnitude for reasonable stellar counterparts
at any given X-ray sensitivity (in larger samples, the nominal $L_X/L_{bol}$-values of some individual stars may exceed $\log L_X/L_{bol}\approx-3$, but excursions toward significantly higher values are very rare). Figure~\ref{fig:limG}
shows the {\it Gaia} magnitude that a star emitting X-rays at the saturation limit will have 
for the sensitivity of eFEDS, which is also approximately the limit for the 
eROSITA survey after eight passes \citep[eRASS:8, see][]{Predehl_2021}.
In particular, virtually
all stellar counterparts of eROSITA sources (eFEDS and eRASS) are expected to 
have G-magnitudes 
brighter than the detection limit of {\it Gaia} and, thus, will be eventually 
included in the {\it Gaia} catalog \citep{Gaia_2016}.  The {\it Gaia} catalog has the additional
benefit of being all-sky and future data releases will improve the completeness
to levels sufficient for identifying stars in the full eRASS survey; 
the current data release (EDR3)
is already complete for stars between G=12\,mag and 17\,mag 
\citep{Gaia_2020}. Given these beneficial properties, we chose to use the 
Gaia catalog as our match catalog (and magnitudes are therefore in the 
VEGAMAG system). 

Our identification scheme aims to identify coronal emitters in 
quiescence. Stars occasionally flare, which can elevate the $L_X/L_{bol}$
quite significantly \citep[see ][subm., for examples]{Boller_2021}. Currently,
stars undergoing flares during the eROSITA observations may show, depending on the 
specific star and flare,  $L_X/L_{bol}$-values that are 
so high ($L_X/L_{bol}\gg-3$) that they are deemed 
incompatible with stars  in our schemes and, thus, may be misclassified as 
non-stellar  (associations with $L_X/L_{bol}$-values somewhat above -3 are 
typically still classified as stellar, see Fig.~\ref{fig:FxFg_all}). 
The methods discussed here do not attempt to correct for the effects of flaring, because
the observing sequence 
of the eROSITA all-sky survey differs from that of eFEDS. For eRASS, flares will be 
easily detected in the survey as each object is 
scanned multiple times. Also, we concentrate on coronal X-ray sources and
other galactic X-ray  sources exist, which include genuine stars  (e.g., CVs).
The origin of the X-ray emission in these objects, however, differs from the 
coronal emission seen in ``normal'' stars and they typically  have 
 high fractional X-ray luminosities. Therefore, they are also unlikely to be classified as stellar
with our methods.

\subsection{Input catalogs and data screening}
The eROSITA and {\it Gaia} catalogs contain entries that are very  unlikely to describe
stars detected by eROSITA and we applied a number 
of filter criteria before performing the stellar identification to remove
such catalog entries.

\subsubsection{Stellar candidates in the eROSITA eFEDS catalog}
We used eROSITA detections from the main eFEDS catalog, which 
contains sources detected in the energy range between 0.2 and 2.3\,keV \citep[see][]{Brunner_2021} and applied the following filter.
\begin{enumerate}
  \item No significant spatial extension (\texttt{EXT\_LIKE}$<6$),\\[-0.3cm]
  
  Extended sources are unlikely to represent stars unless they are a
  blend between two (or more) sources (expected to be very rare).\\[-0.2cm]
  
  \item positive \texttt{RADEC\_ERR}.\\[-0.3cm]
  
  There is a small number of eROSITA sources for which the source detection 
  algorithm  failed to calculate reasonable positional uncertainties. Because
  the positional uncertainty is a key value for our matching algorithm,
  we ignored sources with negative or zero \texttt{RADEC\_ERR} entries.
  
\end{enumerate}
Applying these criteria resulted in $N_X=$~27\,369 eROSITA detections, which constitute 
the X-ray input catalog abbreviated with 
$\mathcal{X}$ in the following.

\begin{figure}[t]
  \centering
  \includegraphics[width=0.49\textwidth]{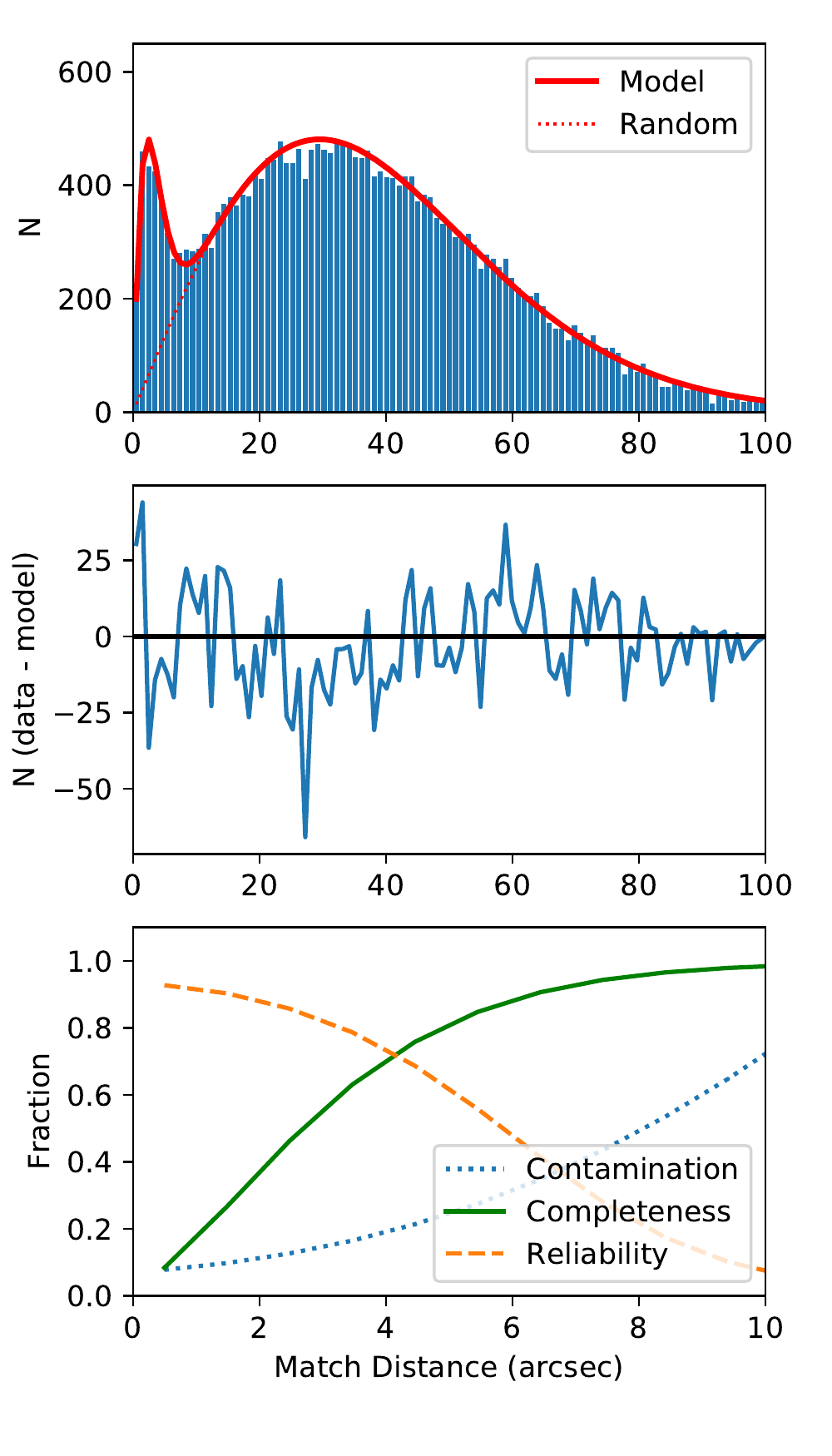}
  \caption{{\bf Top}: Nearest neighbor distance distribution between eROSITA and eligible Gaia sources. Also shown are 
  the best-fit model using Eq.~\ref{eq:model} and the expected number of random matches at each distance. {\bf Middle} Difference between model and data. {\bf Bottom}: Completeness, reliability, and contamination for samples including stars up the given match distance.  
  \label{fig:nearest}}
\end{figure}

\subsubsection{Eligible {\it Gaia} counterparts}
For the {\it Gaia} sources, we  propagated coordinates from the {\it Gaia} EDR3 reference
epoch (2016.0) to the epoch of the eFEDS observation (2019.85). Among the 
{\it Gaia} catalog entries, we 
selected those sources fulfilling the following
conditions to be considered eligible counterparts to eROSITA detected stars.
\begin{enumerate}
\itemsep2pt
  \item Gaia sources above a certain brightness limit given by the X-ray 
      sensitivity, \label{item2}\\[-0.3cm]
      
      The eROSITA X-ray sensitivity limit combined with the range of $L_X/L_{bol}$ ratios
      of stars (see sect.~\ref{sect:intro}) implies that eligible counter parts to eROSITA
      detections must have $G<19$\,mag (for spectral types A to M, see Fig.~\ref{fig:limG}).
      
  \item {\it Gaia} sources must have a three sigma significant parallax 
      measurement (\texttt{parallax\_over\_error}$>3$),\\[-0.3cm]
      
      {\it Gaia} entries that do not have a significant parallax measurement are mostly
      extra galactic or galactic objects so distant ($d \gg 1\,$kpc) that they 
      unlikely represent eligible counter parts to eROSITA detected stars.
      
  \item {\it Gaia} sources that fall into a rather generously defined
      region of the {\it Gaia} color-magnitude diagram (see App.~\ref{sect:isochrones}).\\[-0.3cm]
      
      This region includes
       young (1\,Myr) and old (10\,Gyr) stars with some margin to 
      account for measurement errors. 
      Regions in color-magnitude space where other 
      objects such as cataclysmic variables (CVs) may be found are excluded 
      (below the main sequence).
      We consider the stellar properties for objects outside this region 
      to be too poorly known for our further analysis.
      
\end{enumerate}
We did not filter the {\it Gaia} sources for RUWE, because we found that the sources
with larger RUWE values\footnote{A RUWE value of $1.4$ was suggested for single stars within 100\,pc, 
see \url{http://www.rssd.esa.int/doc_fetch.php?id=3757412}} have 
perfectly acceptable properties and suspect that the selection bias towards 
active stars and binaries or multiples may, at least partially, influence the astrometric solution leading 
to less than perfect RUWE values. In total,  $N_G\approx4\times10^5$ Gaia EDR3 
counterparts fulfill our quality criteria (called $\mathcal{G}$ in the 
following).

\subsection{The catalog fraction \label{sect:CF}}
Among the properties describing any given eROSITA to Gaia association $i \leftrightarrow j$, the 
matching distance is special: 
Measuring the sky positions with very high precision unambiguously informs
us about the correct counterpart, and if such a counterpart does exist
in the match  catalog, because 
sources with too large matching distances have negligible likelihoods to be the 
correct counterpart independent of all other parameters provided that source 
confusion is negligible.

This peculiarity of the sky positions provides a particularly beneficial information
for the matching procedure, namely  an 
estimate for the number of real matches, i.e., the fraction of 
eROSITA sources with a counterpart in the match catalog (the screened Gaia catalog), which we call 
the geometric catalog fraction 
(CF) in the following. The CF can be derived from the measured on-sky 
distances between the eROSITA sources and their nearest Gaia match, because 
the probability distributions to measure a particular match distance is known 
a priori for real and random associations. The distribution 
in match distance depends only on the positional 
uncertainty of the eROSITA source $\sigma_i$ for real associations (in our
context, Gaia sources have negligible positional uncertainties) and on the 
local sky density of eligible sources in the match catalog for random 
associations, respectively. Both properties can be measured independently
of the nature of the eROSITA sources and its membership in the match catalog.

The CF  is a sample 
property affecting all associations equally; it 
cannot be used to select any particular association over another 
association. For example, a CF of 0\,\% 
would imply that even 
near perfect positional matches cannot be considered real while for a 
CF of 100\,\%, large match distances are 
perfectly acceptable, because the correct counterpart must be among the candidates
and one ``just'' needs to
identify the correct counterpart among the ensemble of candidates.

Figure~\ref{fig:nearest} (top) shows the measured nearest neighbor match distance
distribution between eROSITA and Gaia sources for eFEDS. Assuming that the 
positional errors $\sigma_i$ and sky densities $\eta_j$ do not differ between
real and random 
associations, i.e., sample from the same parent population,
the match distance distributions are known for real and random associations and 
the model for the match distance distribution has only one free parameter, namely the CF.
In particular, the match distance distribution is 
\begin{equation}
  d_{real} (r_{ij},\,\sigma_i) = \frac{r_{ij}}{\sigma_i^2} e^{\frac{-r_{ij}^2}{2 \sigma_i^2}} \label{eq:real}
\end{equation}
for real matches  with the match distance $r_{ij}$ for the association between the $i$-th eROSITA 
and the $j$-th Gaia source and the Gaussian positional uncertainty $\sigma_i$, which 
we calculated from \texttt{RADEC\_ERR} as
\begin{equation}
\sigma_i = \sqrt{\frac{\mathrm{RADEC\_ERR}^2_i + 0.7^2}{2}}
\end{equation}
including a systematic uncertainty of 0.7\,arcsec \citep{Brunner_2021}.
The likelihood to find a random association $i\leftrightarrow j$ within $r_{ij}$ 
and $r_{ij}+\,dr$ 
is  proportional to $r_{ij}$ and 
the local sky density. 
Specifically, the distribution of match distances towards
the nearest random neighbor is described by
\begin{equation}
 d_{random} (r_{ij},\,\eta_j) = 2  \pi  r_{ij}  \eta_j e^{- \pi \eta_j r_{ij}^2}  \label{eq:random}
\end{equation}
where $\eta_j$ is the local sky density, which we measured from the local 
neighborhood of any eligible Gaia source.
The model for the measured match distance distribution is then 
\begin{equation}
\displaystyle 
d_{measured}(r) = \sum_{ij}\left( CF\,d_{real}(r,\,\sigma_i) + (1-CF)\,d_{random}(r,\,\eta_j) \right) \label{eq:model}
\end{equation}
where the summation is over all associations between 
eROSITA and their nearest Gaia match, that is, the CF describes the relative 
normalization between real and random associations.

Figure~\ref{fig:nearest} (top and middle) shows that such a model for the 
match distance distrubution accurately describes the measurements and we 
found a CF of 7.5\,\% for eFEDS corresponding to $\N_EFEDS\pm17$ stellar sources. 
As expected, stars represent only a small fraction of the eROSITA detections
in eFEDS. In addition, a thorough 
statistical treatment in the form of a Bayesian mixture model confirms the above
number (to be presented in Freund et al., in prep.).
The bottom panel of Fig.~\ref{fig:nearest} 
shows that almost all real associations have match distances
of 10\,arcsec or less, which is expected based on a median positional error
$\sigma$ of about 3\,arcsec (as the median corrected \texttt{RADEC\_ERR} is about 4.6\,arcsec).

Knowing the CF implies that we know the number of stars (\N_EFEDS), but not 
which specific eROSITA sources are stellar. Knowing the CF also implies 
that when a set of $N=\N_EFEDS$ eROSITA sources is 
classified as stellar, the number of 
stars misclassified as spurious $N_{missed}$  (type~II error or false 
negative) and the number of sources erroneously classified
as stars $N_{spurious}$ (type~I error or false positive) are equal, i.e., 
completeness and reliability are equal. Using 
 the following definitions 
\begin{eqnarray}
\mathrm{completeness} &=& \frac{N- N_\mathrm{spurious}}{N - N_\mathrm{spurious} +N_\mathrm{missed}} \label{eq:completeness}\\
\mathrm{reliability} &=& \frac{N-N_\mathrm{spurious}}{N}\,, \label{eq:reliability}
\end{eqnarray}
Fig.~\ref{fig:nearest} (bottom) shows that
the completeness and reliability are about 70\,\% 
accepting associations up to a fixed match distance of $d_{max}=4\,$arcsec,
i.e., one would classify the correct number of eROSITA sources as stellar, but
almost every third classification would be erroneous. The task at hand is therefore 
to improve the completeness and reliability of a sample containing the \N_EFEDS 
most likely stellar sources using more information than just the match distance.

\begin{figure}[t]
  \centering
  \includegraphics[width=0.49\textwidth]{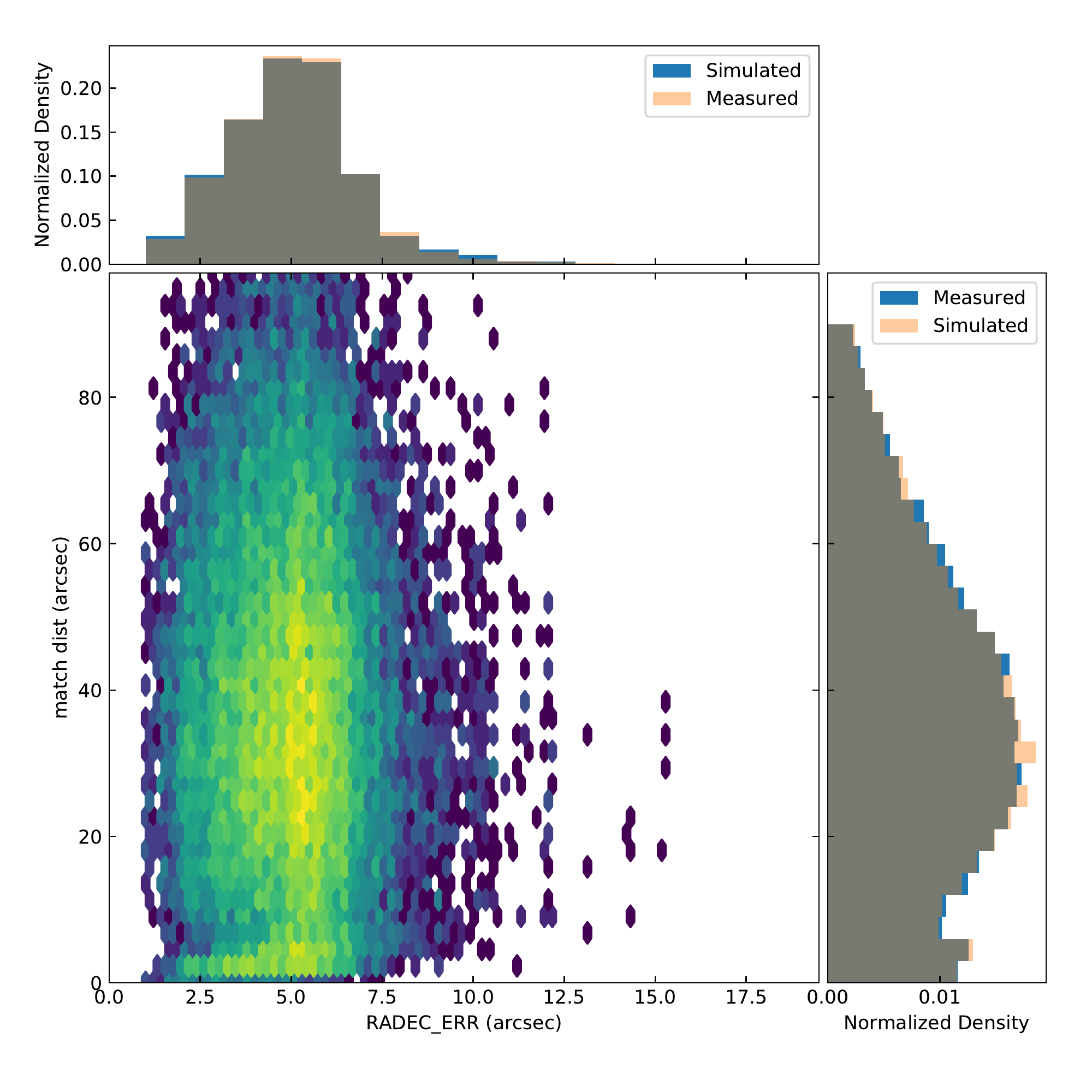}
  \caption{Measured and simulated match distances for eFEDS. In the histograms, 
  the gray bars indicate the overlap between the measured and simulated samples.
  \label{fig:simul_pos}}
\end{figure}

\section{Method I: Support Vector Machine \label{sect:SVM}}
To identify the stellar X-ray sources, we now want to
classify all associations $i\leftrightarrow j$ between eROSITA X-ray sources and 
eligible {\it Gaia} counterparts into the previously defined categories of stellar
and non-stellar associations. This binary classification task resembles problems
regularly approached with ML techniques.

The feature vector, which contains the properties describing an association 
$i\leftrightarrow j$ such as the match distance $r_{ij}$,
plays a crucial role for all ML classification methods,
raising the question which features should be included in it. 
In principle,
any combination of properties available from the Gaia and eROSITA catalogs
may be used to populate the feature vector describing an association.
However, the eFEDS dataset is somewhat limited in source numbers for ML
methods to extract the relevant properties in the feature vector in addition
to fitting or training. Therefore,
we opted for an astrophysically motivated choice of
properties, which are listed in Table~\ref{tab:properties}. Specifically, we
constructed feature vectors with $M_X=2$ X-ray ($F_X$, $\sigma$) and $M_G=4$ Gaia features
($F_G$, $plx$, $\eta$, and $BP-RP$) plus the match distance $r_{ij}$, so that the
feature vector is seven dimensional  ($\mathbf{p}_{ij} \in \mathbb{R}^{M_X + M_G + 1} = \mathbb{R}^7$).

Our choice of features has the highly advantageous property that
the probability for a correct classification
is a monotonic function in many features listed 
in Table~\ref{tab:properties}. For example,
an association $i\leftrightarrow j_1$ with a Gaia source $j_1$ being less distant (in pc)
is generally the more likely counterpart than a Gaia source $j_2$ at a larger distance 
when all other features are equal, because the number of chance alignments 
increases with distance. Similarly,
the higher the optical flux, the higher is the likelihood that the association under 
consideration is correct, because optically bright sources are rare and, clearly,
the smaller the on-sky match distance, the higher is the likelihood of a 
correct association. In fact, we used the latter property to strongly reduce the number of 
to-be-classified associations by considering only plausible ones with match 
distances of up to 60\,arcsec.

The monotonic behavior of the feature vector with respect to the association 
probability makes our classification task ideally suited for a support 
vector machine (SVM) approach \citep{Cortes_1995}. A SVM 
classifies a feature vector $\mathbf{p}_{ij}$, 
which in our case characterizes
an association, into the two categories of stellar and non-stellar associations.
The basic idea of a SVM classifier (SVC) is to use a training sample with labeled data to 
construct a so-called maximum-margin hyperplane separating the two 
categories in feature space. The confidence in the classification increases 
with distance from this hyperplane; samples near the hyperplane have less secure 
classifications compared to samples far away (in feature space) from the 
hyperplane. This behavior is well matched to the content of our feature vector
and we used the SVC implementation 
of scikit-learn \citep{scikit-learn} with a polynomial kernel 
to categorize associations $i \leftrightarrow j$ in the following.

\subsection{Training and validation Samples \label{sect:SCV_training_sample}}
The quality of the training sample is crucial for the final classification.
Among the available information on the associations,
the match distance stands out for indicating likely matches
independent of any other property. Therefore,
we constructed our final training sample in two steps, first, focusing on
geometry and, second, incorporating physical properties.

\subsubsection{Selection of the training sample with a geometric SVC}
In a first step, we constructed a geometric training sample based on the best positional
associations and some quality criterium as described in the following.
The reliability of an association depends on the positional uncertainty of the
X-ray source $\sigma$ and on the local density of eligible counterparts $\eta$, because
the expected number of random matches scales as $\eta \sigma^2$, which must be small for
a reliable association.
Hence, the same match distance may be perfectly acceptable 
for low sky densities while unacceptable for high sky densities.

To identify the $N$ best geometric associations for the final training sample,
we used a geometric SVC with a feature vector consisting of
match distance, positional uncertainty, and local sky density.
As the geometric distributions for real and random source are known (cf. sect.~\ref{sect:CF})
as well as the CF, sampling from
these distributions provided us with large ($N\gtrsim10^4$) geometric training sets. 
These training data reproduce the geometric properties of the eFEDS field, and here we also have
the labels to train the SVC. As an example, Fig.~\ref{fig:simul_pos} shows 
the resulting match distance distribution for the geometric training sample, demonstrating
the excellent overlap with the observed distribution in the data.
The geometric SVC was then trained to classify 
associations based on geometry allowing a 
contamination of 5\,\% spurious associations,
which we found to balance sample size and contamination.

\begin{figure}[t]
	\resizebox{\hsize}{!}{\includegraphics{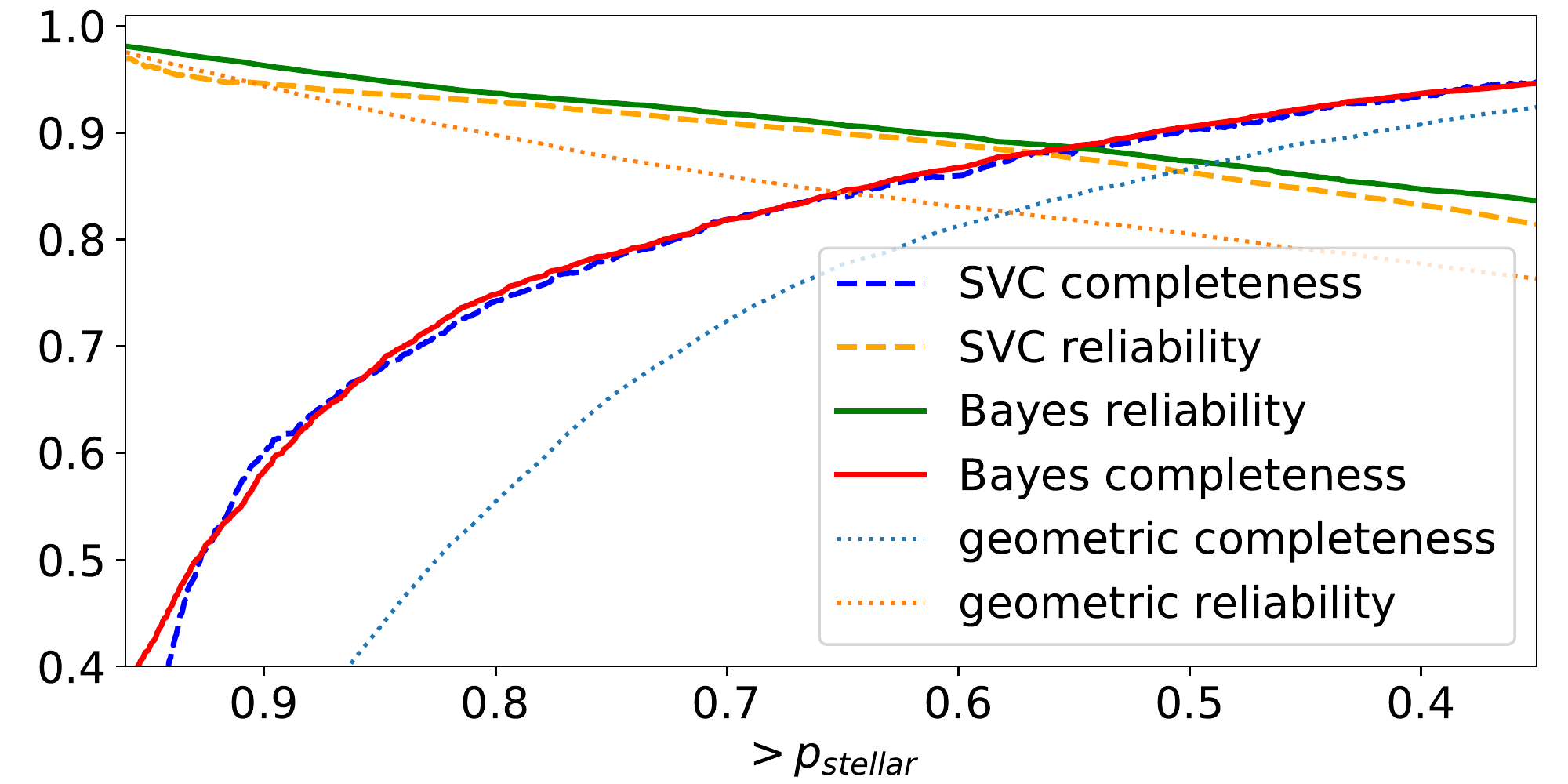}}
	\caption{Completeness and reliability of the stellar identifications for the eFEDS sources as a function of the probability cutoff for the SVM and Bayesian algorithms as well as the purely geometric version.}
	\label{fig: completness and reliability}
\end{figure}

This geometric SVC was then used on the real data classifying 627 associations
as real. As a cross-check, we applied the geometric SVC to
randomly shifted eFEDS X-ray sources, which produces only random matches by construction.
In this case, the geometric SVC classified 36 associations as stellar, which
corresponds to an empirical contamination level of just under 6\,\% well in line with the 
expected value. 

In a second step, we applied a physical screening of the geometric training sample.
In particular, we excluded associations with counterparts yielding
\hbox{$L_X > 2\cdot 10^{31}$~erg\,s$^{-1}$} and 
empirically defined a limit of $\log(F_X/F_G) < (BP-RP)\times 0.655 - 3.22$
for the fractional X-ray fluxes of the training sample objects,
larger $F_X/F_G$-values are well above the saturation limit and unlikely to be produced by stellar coronal X-ray emission in quiescence; only highly active A-type stars may lie above the limit, but are not present in our data anyway.
We further excluded a few associations with counterparts brighter than $G=5.5$~mag, 
because they potentially cause optical loading (although the association itself
may still be the correct one, just the properties could be skewed).
Finally, we identified 537 bona fide stellar associations to represent
the physical training sample. 

This physical training sample contains, by construction, only associations with relatively
small match distances, which do not sample
the ensemble of expected match distances of true associations well.
Therefore, we constructed the final training sample by reasigning new match distances sampling 
the expected distrubution for real associations, just like for the 
geometric training sample.
This time, however, we kept the X-ray and optical properties of the previously 
identified associations in the respective feature vectors.
The final training sample also
includes random associations, which were selected in proportion to 
the ratio implied by the CF, i.e, the training set reproduces the true ratio between
real and spurious associations. The properties of random associations can easily be explored
by shifting the eROSITA X-ray sources w.r.t. the background of eligible Gaia counterparts,
which is what we did to incorporate them into the final training sample.

\begin{table*}[t!]
  \caption{Properties used for classification \label{tab:properties}}
  \begin{tabular}{clccc}
  \hline
  \hline
  & Name & Abbreviation & Description & Unit \\
 & & in text & \\ 
  \hline
  1 & Angular separation & $r_{ij}$ & Angular distance between eROSITA and Gaia source & arcsec\\
  2 & Local sky density & $\eta$ & Local sky density of eligible Gaia sources & arcmin$^{-2}$\\
  3 & Positional uncertainty & $\sigma$ & Uncertainty in position for eROSITA source & arcsec \\
  4 & Optical flux & $F_G$ & Optical flux in Gaia band & erg\,s$^{-1}\,$cm$^{-2}$ \\
  5 & X-ray flux & $F_X$ & X-ray flux for eROSITA source & erg\,s$^{-1}\,$cm$^{-2}$\\
  6 & Gaia color & BP-RP & Color of Gaia source & mag \\
  7 & Distance & $d$ & Distance of Gaia source (from parallax) & pc\\
  \hline
  \end{tabular}
\end{table*}

\subsubsection{Validation sample}
The validation sample was constructed similarly as the random associations in the 
training set (see above), that is, by matching randomly shifted  eFEDS sources. 
This validation sample contains \emph{no} correct associations and all
associations classified as stellar by the SVM must be spurious (false positives)
by construction.
Usually, this
would give only part of the desired information, i.e., would not allow an
assessment of the completeness. Having an accurate estimate for the
number of stellar eFEDs sources from the CF, however, provides this missing piece of information and the 
completeness is simply
\begin{equation}
\mathrm{Completeness} = \frac{N-N_{spurious}}{N_{stars}}\,,
\end{equation}
where $N$ is the number of eFEDS sources with at least one 
association classified as stellar
and $N_{stars}=\N_EFEDS$ is the 
known number of stars from sect.~\ref{sect:CF}. 
Therefore, we do not need a validation sample containing real stellar X-ray 
emitters to evaluate the completeness and reliability of the classifier.

\subsection{Preprocessing}
Scales and ranges differ between the features used for the classification,
e.g., X-ray fluxes are roughly in the range of 
$10^{-14}$\,erg\,s$^{-1}$\,cm$^{-2}$ to  $10^{-12}$\,erg\,s$^{-1}$\,cm$^{-2}$ 
while we considered match distances between zero and 60\,arcsec. In such cases, 
it is often recommended to scale each feature
to some ``standardized'' distribution, for example, to normalize to zero mean and 
standard deviation one. Empirically, however, we found that this does not 
provide good results for the problem at hand in terms of sample reliability 
and reliability. Therefore, we opted for individually 
scaling the features such that 
numerical values are roughly in the range between 0 and 10 
and, e.g., used logarithmic fluxes. The exact scaling values
and zero points impact the importance of individual features in the 
mixed terms of the polynomial kernel and may help, to some degree, to
adjust their respective weights in the SVM.

\subsection{Optimization goal \label{sect:optimization}}
Classification tasks often imply a tradeoff between
completeness and reliability, e.g., one may want to have a clean sample
with little contamination or a sample that captures the largest number 
of objects at the expense of a larger contamination level. 
In contrast to many other classification tasks, we have a good estimate of  the 
correct number of stars---just not which \emph{individual} eROSITA sources
are the stars.
Therefore, we chose to optimize the classifier such that the correct
number of sources are classified as stars, that is, have at least one 
likely stellar association.
This choice implies that the 
number of stars misclassified as non-stellar $N_{missed}$
and the number of sources erroneously classified
as stars $N_{spurious}$  are equal. 
However, this also implies that
we do not necessarily achieve 
the highest possible accuracy; it is conceivable that a solution exists with
a larger number of correctly classified objects at the cost
of, e.g., a larger imbalance between $N_{missed}$ and $N_{spurious}$. 
Inspecting the behavior of the SVC,
we found that the hyper-parameters resulting in $N_{missed}=N_{spurious}$ 
span a well-defined 
path in the parameter space. From those solutions, we chose
the parameters that achieve the highest accuracy.

\begin{figure}[t]
	\resizebox{\hsize}{!}{\includegraphics{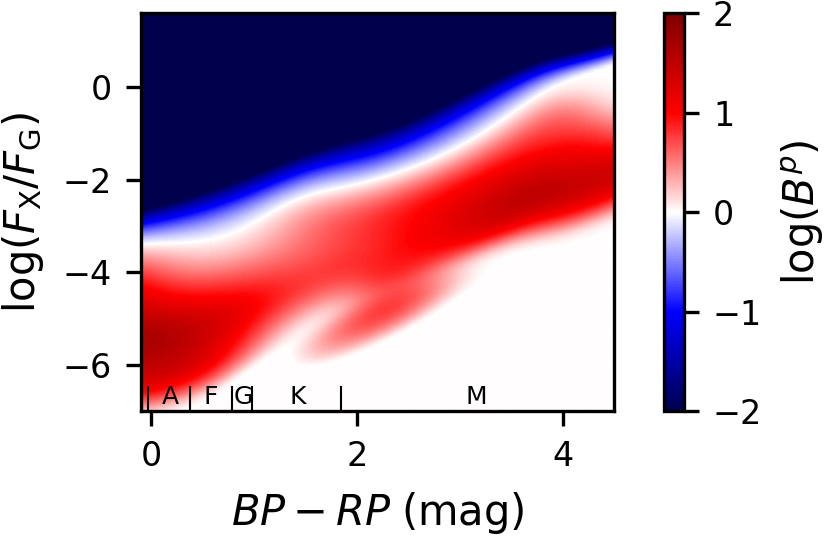}}
	\caption{Distribution of the Bayes factor $B^p$ as a function of the activity $F_X/F_G$ and the color $BP-RP$ for the eFEDS counterparts. The ranges of the spectral types are indicated at the bottom of the figure.}
	\label{fig: Bayes factor}
\end{figure}

\subsection{SVC results \label{sect:SVC_results}}
We used a third degree polynomial kernel and the properties from 
Tab.~\ref{tab:properties}  replacing the X-ray and optical fluxes with 
their flux ratio $F_x/F_G$ resulting in a six dimensional feature vector. The
SVC was then trained with the above described training sample and the 
hyper-parameters adjusted so that $N=\N_EFEDS$ eFEDS sources are classified 
as stellar. 
With this requirement, the expected number
of eROSITA sources randomly classified as stellar within the \N_EFEDS sources
is $N_{spurious} = 239$ so that also $N_{missed} = 239$ real sources are not classified as stellar by the SVC.
This corresponds to 
a completeness and reliability of 88.4\,\% (cf. Eqs.~\ref{eq:completeness} and \ref{eq:reliability}
for the definitions of completeness and reliability). The resulting sample properties
will be discussed in sect.~\ref{sect:comparison} together with the sample resulting 
from the Bayesian approach.

\begin{figure*}[t]
  \centering
  \includegraphics[width=0.89\textwidth]{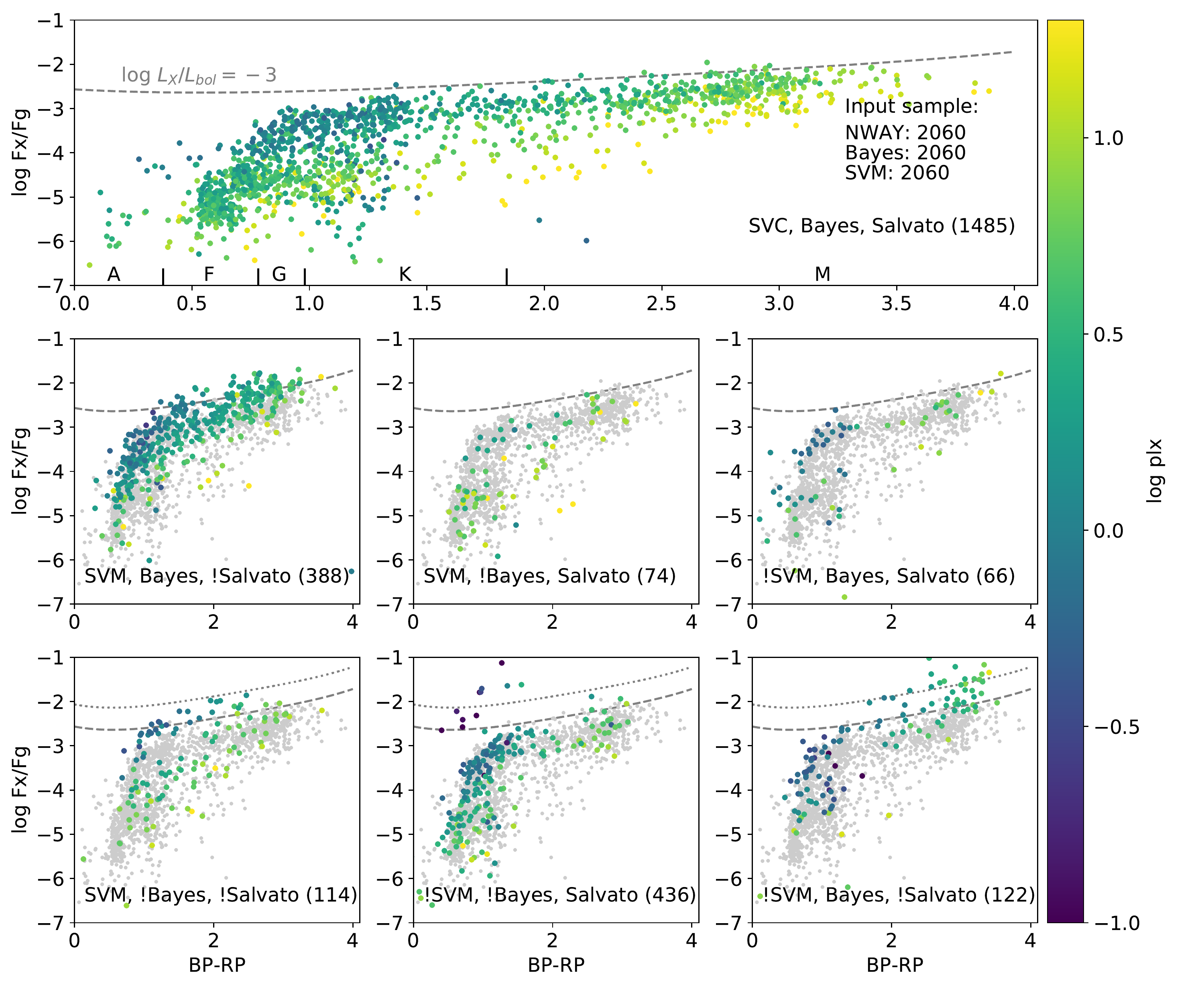}
  \caption{Ratio between X-ray and G-band fluxes as a function of the associated Gaia 
  object's BP-RP color. The objects are colored according to the parallax. The 
  label in the lower part of all seven panels indicate in which sample
  the colored objects belong: SVM, Bayes, and Salvato describe the method and  ``!'' equals 
  \emph{not}, that is, that the colored identification are not identified by the method that is 
  preceded by the ``!''. The number in bracket indicates the respective number of sources.
  In rows two and three, the gray dots represent the objects of the top panel for reference.
  The dotted line in the bottom row indicates $\log L_X/L_{bol} = -2.5$.
  \label{fig:FxFg_all}}
\end{figure*}

The SVC does not directly provide well calibrated probabilities $p_{ij}$ for 
the associations $i \leftrightarrow j$.
Therefore, one cannot easily construct samples with different completeness or reliability
values applying certain cuts in association probability $p_{ij}$; rather a new training 
run would be required to
achieve the best performance for a specific completeness or 
reliability level. However, the number of random 
associations for a certain cutoff in $p_{ij}$ can be directly 
derived by applying the algorithm 
to the validation sample. A natural choice is therefore to empirically calibrate the association 
probabilities $p_{ij}$ such that 
\begin{equation}
N_{spurious}(p>p_{min}) = \sum_{p>p_{min}} 1 - p_{ij}
\end{equation}
using the validation sample. 
Figure~\ref{fig: completness and reliability} shows the resulting 
completeness and 
reliability levels as a function of the cutoff value $p_{min}$ in the thus calibrated 
association probabilities $p_{ij}$. Here, we used the previous
definitions for completeness and reliability replacing $N$ with $N_>(p_{min})$, i.e., 
the number of associations above the cutoff-threshold.

\begin{figure*}[t]
  \sidecaption
  \includegraphics[width=12cm]{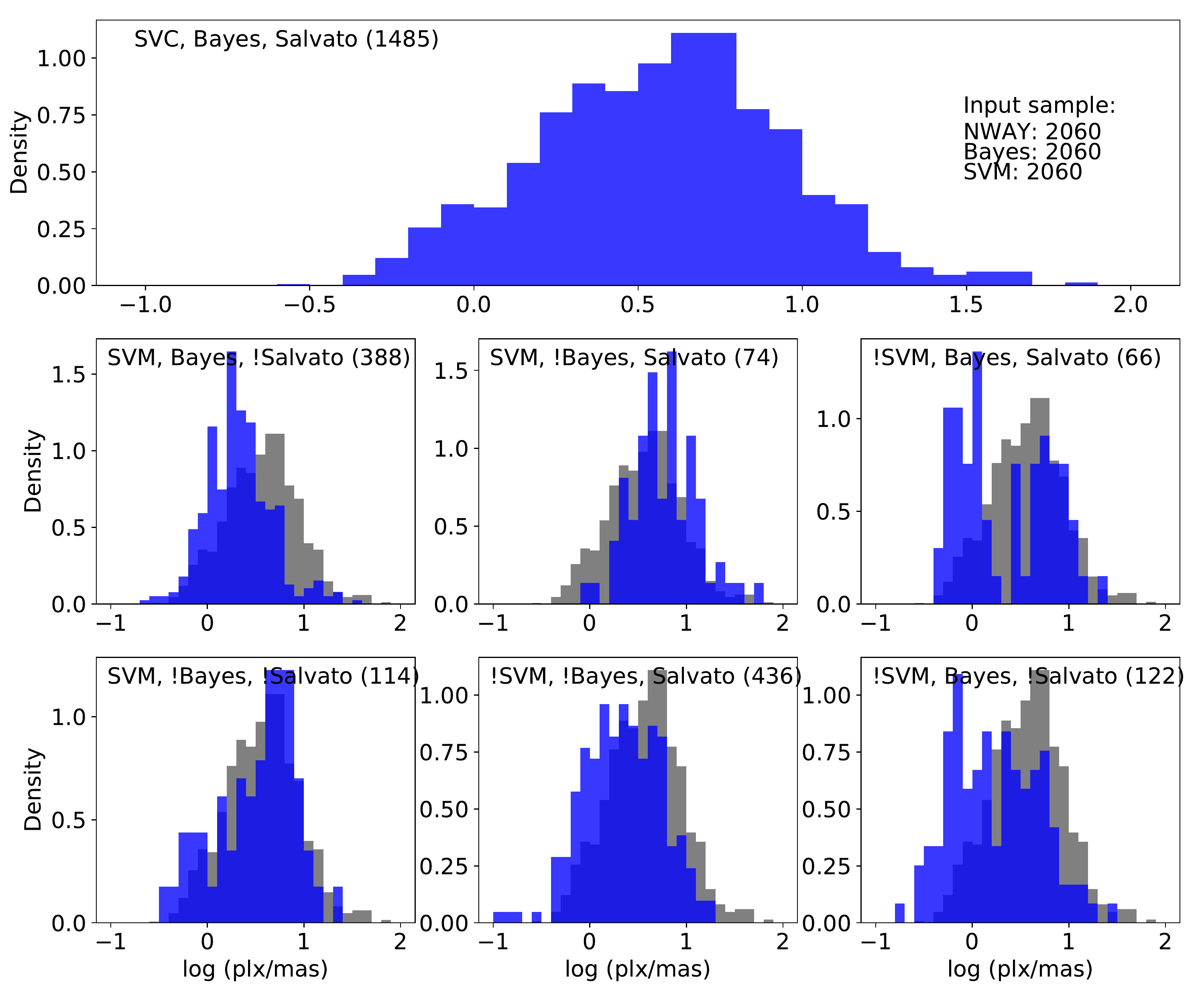}
  \caption{Histograms for the parallaxes of the identified Gaia counterparts. Individual panels as in Fig.~\ref{fig:FxFg_all}. The parallax distribution of the associations identified by all three methods (top panel) is shown in the background  as the gray histogram 
  in the middle and bottom rows.
  \label{fig:log_plx}}
\end{figure*}

\section{Method II: Bayesian approach} \label{sect:Bayesian}
In our Bayesian matching framework, the prior probability of picking by chance 
the correct counterpart is updated after obtaining data of the source position 
and properties. In that sense, we followed similar Bayesian catalog matching techniques 
described by \citet{Budavari_2008} and successfully implemented in 
cross-matching tools such as NWAY \citep{Salvato_2018}.

\subsection{Distance based matching probability}
\label{sec: distance based matching probability}
Again, we consider the problem of matching $N_G$ eligible stellar candidates to $N_X$ X-ray sources. Therefore,
we define the following hypotheses, the probabilities of which we want to compare:
\begin{table}[H]
	\begin{tabular}{llp{7cm}}
		$H_{i1}$ &:& the i-th X-ray source is associated with the $j=1$ counterpart \\
		\multicolumn{3}{c}{$\vdots$} \\
		$H_{ij}$ &:& the i-th X-ray source is associated with the j-th counterpart \\
		$H_{i0}$ &:& the i-th X-ray source is not associated with any of the counterparts. \\
	\end{tabular}
\end{table} 

Extending upon the scheme of \citet{Budavari_2008}, we
know that only a fraction CF of the X-ray sources can be associated with 
one of the $N_G$ counterparts based on the distribution of the 
match distances $r_{ij}$ and sky densities $\eta_j$  (cf. Sect.~\ref{sect:CF}).
Hence, we derived the prior probability that none of the optical counterparts is associated with the i-th X-ray source through  
\begin{equation}
P(H_{i0}) = 1-\mathrm{CF}, 
\label{equ: prior no match}
\end{equation}
and the prior probability that the j-th counterpart is the identification of the i-th X-ray source by 
\begin{equation}
P(H_{ij}) = \frac{\mathrm{CF}}{N_G} = \frac{\mathrm{CF}}{\eta\Omega}, 
\label{equ: prior match}
\end{equation}
where $\eta$ is the source density in the area of the sky given by $\Omega$.

Next, we considered the positions of the X-ray source and the eligible stellar candidates.  Neglecting the positional uncertainty of the stellar candidates, the likelihood of obtaining the data $D_i$ given that the j-th counterpart is the correct identification of the i-th X-ray source was estimated through
\begin{equation}
P(D_i|H_{ij}) = \frac{1}{2\pi\sigma_i^2}e^{-\frac{r_{ij}^2}{2\sigma_i^2}} \; ,
\label{equ: likelihood match}
\end{equation}
where $\sigma_i$ is the positional uncertainty of the i-th X-ray source and $r_{ij}$ is the angular separation between the i-th X-ray source and the j-th counterpart (cf. Eq.~\ref{eq:real}). Counterparts for which \hbox{$r_{ij}\gg\sigma_i$} can be neglected
based on the exponential term.
Assuming the X-ray source is not associated with any of the counterparts, the likelihood of the data becomes
\begin{equation}
P(D_i|H_{i0}) = \frac{1}{4\pi}\; .
\end{equation}
In accordance with \citet{Budavari_2008},
we then obtained the geometric Bayes factor
\begin{equation}
B_{ij}^g = \frac{P(D_{i}|H_{ij})}{P(D_{i}|H_{i0})} = 
\frac{2}{\sigma_i^2}e^{-\frac{r_{ij}^2}{2\sigma_i^2}} \ .
\label{equ: Bayes factor}
\end{equation}

Applying Bayes' theorem, we computed the posterior probability that the j-th counterpart is the correct identification through 
\begin{equation}
p_{ij} = P(H_{ij}|D_i) = \frac{P(D_i|H_{ij})\cdot P(H_{ij})}{\sum_{k=0}^{N_G}P(D_i|H_{ik})\cdot P(H_{ik})} \; . 
\label{equ: posterior matching probability}
\end{equation}
The probability that any of the stellar counterparts is the correct identification, which is equivalent to
the X-ray source being stellar given a complete and uncontaminated counterpart catalog of all and only stars, is given by
\begin{equation}
p_\mathrm{stellar} = \frac{\sum_{k=1}^{N_G}P(D_i|H_{ik})\cdot P(H_{ik})}{\sum_{k=0}^{N_G}P(D_i|H_{ik})\cdot P(H_{ik})}. 
\label{equ: posterior stellar probability}
\end{equation}

\subsection{Consideration of additional source properties}
\label{sec: considering of additional source properties}
Additional properties of the counterparts and X-ray sources can be considered to identify the best match. For example, the matching probability can be increased if the activity estimated by the X-ray flux and the optical brightness of the counterpart meets the expectations of a stellar source and few random associations are expected at such activity levels.
Technically, this was achieved by expanding the geometric Bayes factor by another factor $B_{ij}^p$, which
represents the expected ratio between physical associations and random associations, so that
\begin{equation}
B_{ij} = B_{ij}^g \times B_{ij}^p \; .
\label{equ: likelihood match activity}
\end{equation}
As additional properties, we used
the X-ray to optical flux ratio, $F_X/F_G$, and the $BP-RP$ color.

\begin{figure*}[t]
  \sidecaption
  \includegraphics[width=12cm]{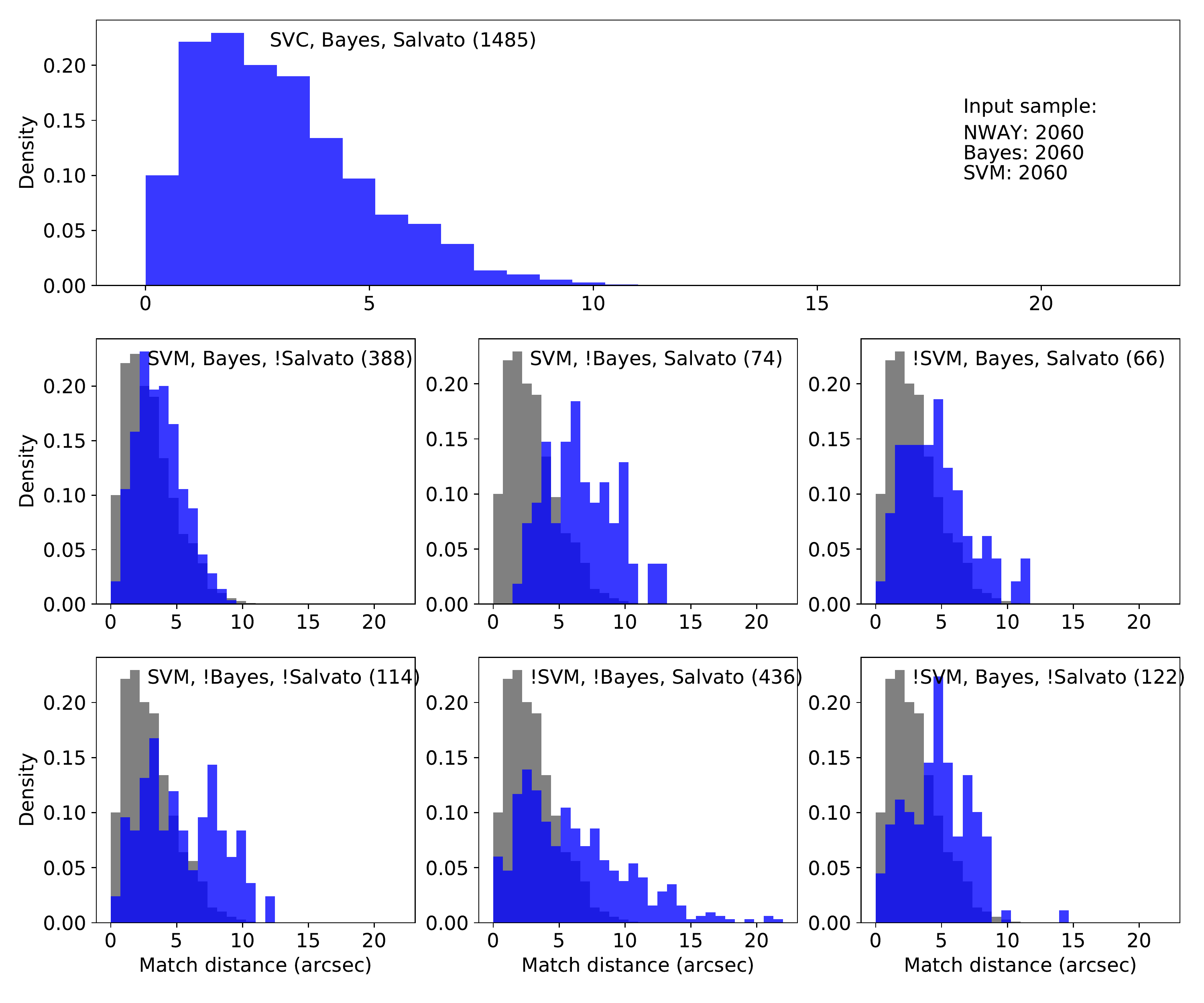}
  \caption{Histograms for the match distance to the associated Gaia counterparts. Similar to Fig.~\ref{fig:log_plx}.
  \label{fig:match_dist}}
\end{figure*}

We estimated the factor $B_{ij}^p$ from the data themselves. In particular, we constructed a training set of highly probable geometric counterparts. To obtain a clean training set, we selected the counterparts with a posterior geometric matching probability $>0.9$ only (cf. Eq.~\ref{equ: posterior matching probability}). Of the thus identified 577 reliable counterparts, we expect 32 to still be spurious. Applying the same screening procedures as for the SVC training sample (cf. Sect.~\ref{sect:SCV_training_sample}), we arrived at a training set with 494 bona fide coronal sources. To derive the distribution of specific properties for spurious association, we again shifted the X-ray sources
arbitrarily on the fixed background of counterparts in the eFEDS field and studied the thus constructed set of explicitly
random matches.

As our training sample remains rather small,
some regions in the $F_X/F_G$ versus $BP-RP$ plane remain sparsely populated, which confounds the estimation of $B_{ij}^p$. This affects, e.g., M-type sources at low $F_X/F_G$ values, which are reasonable X-ray emitters by physical standards, but are unlikely to be detected in X-rays at the eFEDS sensitivity. Therefore, we adopted a value of one for $B_{ij}^p$ in this region of inactive M-dwarfs. On the other end of the main sequence, early A- and B-type stars with high $F_X/F_G$ values can be excluded as true identifications based on physical grounds, here $B_{ij}^p$ goes to zero, which is tantamount to assuming that essentially all such matches are random associations. 
Nonetheless, for most of the counterparts, the details of the estimation of $B_{ij}^p$ have minor impact on the derived matching probabilities.

In Fig.~\ref{fig: Bayes factor} we show the resulting map of Bayes factors $B_{ij}^p$ for the eFEDS counterparts. For example, F-type sources with an activity level of around $\log(F_X/F_G) \approx -5$ are weighted up, while counterparts with $\log(F_X/F_G) > -2$ are weighted down by considering the additional properties.  The Bayes map generally appears rather smooth. However, regions of the Bayes map sparsely populated by training and control sources may show some distinct structure, in particular the increase of the Bayes factor for sources around $F_X/F_G = -5$ and $BP-RP = 2.2$~mag. The influence of these ``low source density regions'' on our identifications is marginal because  the number of counterparts in these regions is also very small and in fact,  neither the SVM nor the Salvato~et al. method show an excess or deficit of sources in  this part of the diagram when compared to the Bayesian method (see Fig.~\ref{fig:FxFg_all}).
We also note that although sources with high X-ray to optical flux ratios are excluded from the training set, such sources can nevertheless be identified in the final sample if their positional match is good or the number of expected spurious association with such properties is small because then the Bayes factor $B_{ij}^p$ is still about unity.

\subsection{Bayesian results}
Applying the matching procedure described in Sects.~\ref{sec: distance based matching probability} and \ref{sec: considering of additional source properties}, we obtained a stellar probability for every eFEDS source. In contrast to the SVC, these probabilities are well calibrated so that the completeness and reliability of a sample selected by a specific probability threshold can be derived directly. The expected number of missed (false negatives) and spurious (false positives) stellar identification is estimated through 
\begin{eqnarray}
N_\mathrm{missed} &=& \sum_{N_<} p_\mathrm{stellar, <}, 
\label{equ: num missed theo} \\
N_\mathrm{spurious} &=& \sum_{N_>} 1-p_\mathrm{stellar, >}
\label{equ: num spurious theo}
\end{eqnarray} 
where $N_>$ and $N_<$ are the number
of sources above and below the threshold and their probabilities are denoted by $p_\mathrm{stellar, >}$ and $p_\mathrm{stellar, <}$, respectively. Completeness and reliability of the obtained sample were estimated through Eqs.~\ref{eq:completeness} and \ref{eq:reliability} 
by replacing $N$ with $N_>$.

In Fig.~\ref{fig: completness and reliability} we present the expected completeness and reliability obtained with the Bayesian approach as a function of the stellar probability cutoff. At $p_\mathrm{stellar} \approx 0.58$ the expected number of stellar sources in the eFEDS field is recovered, here, about 11~\% of the identifications are expected to be spurious and the same fraction of stellar eFEDS sources is expected to be missed. Going to larger stellar probabilities, the completeness decreases and the reliability increases, as expected.
We note that these values were empirically verified by applying our Bayesian identification procedure to arbitrarily shifted eFEDS sources similar to the SVC test sample.

\begin{figure*}[t]
  \sidecaption
  \includegraphics[width=12.5cm]{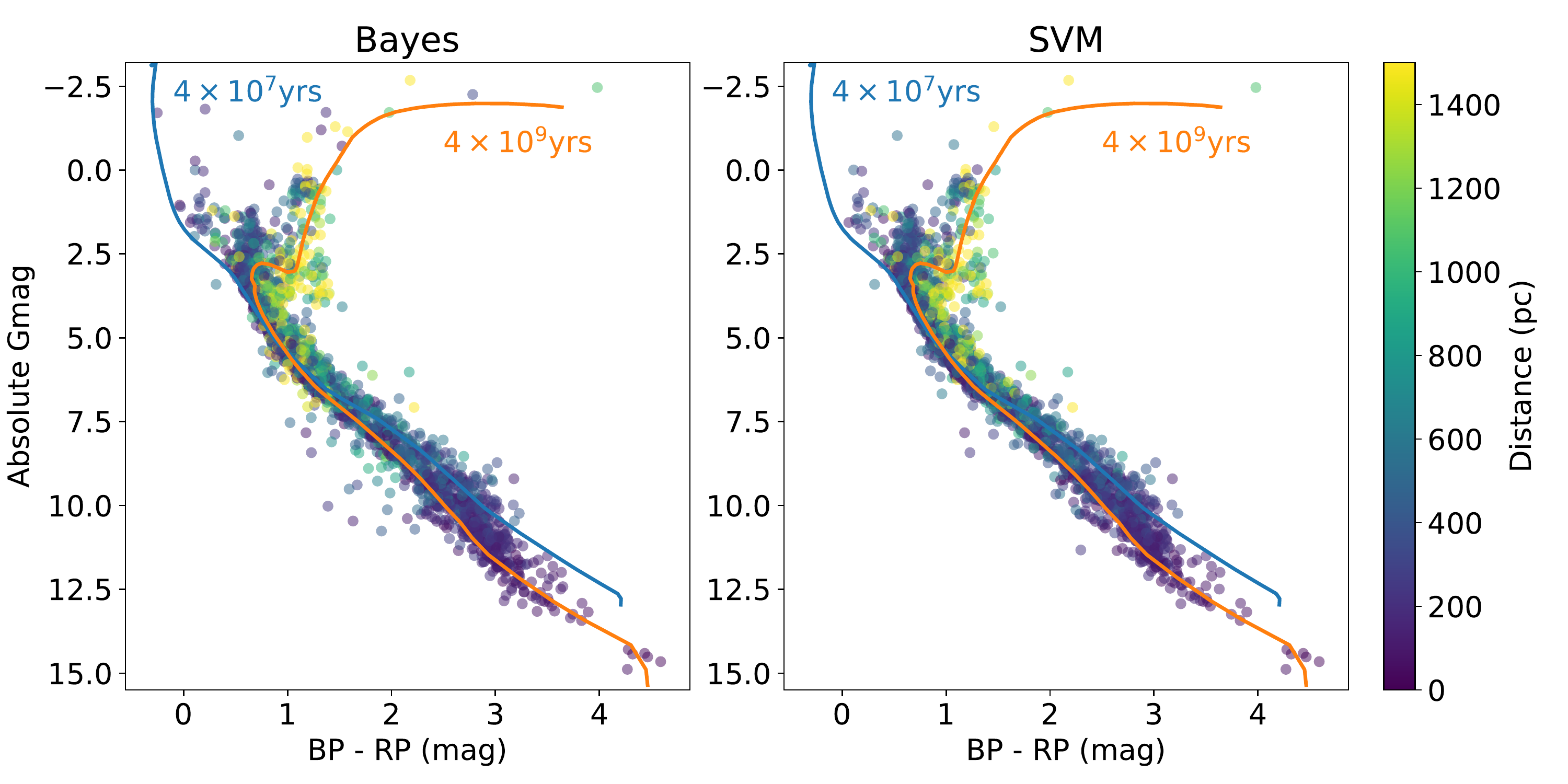}
  \caption{Color-magnitude diagrams for the identified stellar sources ({\bf Left}: Bayes, {\bf right}: SVM). The color indicates the distance of the sources. Isochrones are shown for two representative stellar ages of $4\times10^7$ and $4\times10^9$ years using the PARSEC isochrones \citep{Bressan_2012}.
  \label{fig:HRD}}
\end{figure*}

\section{Results and comparison between approaches\label{sect:comparison}}
We are interested in the physical properties of the stars detected by eROSITA,
more precisely, in the resulting association properties. Therefore,  
we base our comparison on associations and not on the classification of an eROSITA source 
as stellar; a discussion of the overlap in stellar classifications is provided 
in \citet{Salvato_2021}.

\subsection{Comparison samples}
To compare the samples identified by the different methods, we cut the SVM 
and Bayesian catalogs in $p_{ij}$ such that 
\N_EFEDS eROSITA sources are classified as stellar.
We also include the catalog presented  in \citet{Salvato_2021}, which provides
counterparts to all eFEDS sources regardless of their galactic or extra galactic nature. These authors
used  a training sample of 23\,000 {\it XMM-Newton} and {\it Chandra} sources with secure counterpats 
and the final counterpart identification is done after
comparing the associations obtained with NWAY \citep{Salvato_2018} with a
modified version of Maximum Likelihood \citep{Sutherland_Saunders1992}
as described in \citet{Ruiz18}. Again, we cut their sample of stellar 
sources in $p_{any}$ to obtain 2060 stellar eROSITA sources. Furthermore, 
we mapped their Gaia DR2 IDs to EDR3 IDs for the comparison noting 
that not all DR2 sources could be unambiguously mapped to a EDR3 source.

The SVM and Bayesian approaches perform about equally well with an overlap of 
1873 associations in both samples 
(91\,\%). This fraction incidentally coincides with the expected 
reliability of the respective samples. Therefore, it is possible  albeit unlikely  that all true associations
are among the shared associations and that all spurious associations are being identified by one method alone. The resulting sample properties in terms of completeness and reliability are very similar for the SVM and Bayesian approaches (see Fig.~\ref{fig: completness and reliability}). The differences are compatible with the Poisson noise for the number of missed or spurious sources at the respective \hbox{$p_{stallar}$-thresholds}. 

The overlap between the results by \citet{Salvato_2021} and the SVM and Bayesian approaches
is smaller with about 1550 and 1558 associations in common, respectively.
This smaller overlap is partly caused by the different Gaia data releases 
(145 associations are affected) and 
by the construction of the sample since we cut the sample to the \N_EFEDS ``best''
associations. The overlap in stellar classifications of eROSITA sources
is much higher \citep[cf.][]{Salvato_2021}.

\subsection{Sample properties}
We show the physical properties of the resulting samples in Figs~\ref{fig:FxFg_all}
to \ref{fig:HRD}. Figures~\ref{fig:FxFg_all} to \ref{fig:match_dist} display 
different properties but their structure is identical. They all 
contain seven panels organized in three rows: The first row shows the properties of the
associations overlapping between all 
three methods (1484 in total). The second row shows the properties of the 
associations 
overlapping between only two methods and we present one panel for each of the 
possible combinations. For each panel, we provide an annotation indicating
which sample(s) contain the displayed associations. We use a ``\texttt{!}''
to indicate that the associations are not in the 
sample following the ``\texttt{!}'', for example, the left panel in the second row with the 
\texttt{SVM, Bayes, !Salvato} annotation implies that the 
associations are in the SVM and Bayesian samples but not among the \N_EFEDS 
best \citet{Salvato_2021} stellar associations.
Finally, the third row shows the associations exclusively
contained in only one sample. To ease the comparison between the different 
samples, we display in rows two and three of Figs.~\ref{fig:FxFg_all} to 
\ref{fig:match_dist} the  associations overlapping between all three methods
(shown in the top panel of each figure) in the background (in gray). Numbers in brackets
following the sample description indicate the number of associations (displayed in color).

Figure~\ref{fig:FxFg_all} shows that the $F_X/F_G$ distributions are relatively similar for all identification methods 
Noticeable is that all methods classify some associations above the 
saturation limit of $\log L_X/L_{bol}=-3$ as stellar (gray dashed line in Fig.~\ref{fig:FxFg_all}).
These associations between an eROSITA source
and a Gaia source have a combination
of high positional match likelihoods and are, for SVC, relatively nearby, which
counterbalances the high $F_X/F_G$-values and lead to the stellar classification. At least 
some cases of  unusually high $F_X/F_G$ values result from flaring: 65 variable sources are 
discussed in Boller et al., this issue. Most of these variable sources are associated with stars and show, when sufficient X-ray coverage is available, that the X-ray light curves resemble the typical shape of flaring due to magnetic activity. Furthermore, most sources are within
0.5\,dex of the boundary for all methods (dashed line in Fig.~\ref{fig:FxFg_all}). 
The methods employed by \citet{Salvato_2021} identify a small number of sources ($\sim10$) with Gaia sources leading to 
quite high $F_X/F_G$ values, probably  because their association is based on
charaterizing the entire SED. In total, the number of sources above the 
saturation limit remains small: neither method associates more than hundred eFEDS sources 
with a Gaia source leading to $F_{X}/F_{G}$ values implying X-ray emission above the 
saturation limit ($<5$\,\% of the stellar classifications).

Overall the associations classified as stellar fall within the expected BP-RP-color vs $F_X/F_G$ range of 
stellar sources for all three methods. The main features are (a) high X-ray activity levels for M~dwarfs
close to the saturation level, (b) a substantial spread in X-ray activity for stars of spectral types 
F to early K, and (c) the onset of magnetic activity at late A/early F spectral types. These properties
are also seen for the sample of stellar X-ray sources in the XMM-Newton slew survey catalog identified
by \citet{Freund_2018} and may be a general property of flux limited X-ray surveys.

We show in Fig.~\ref{fig:HRD} the positions of the Gaia associated with the stellar eROSITA in the color-magnitude diagram. By construction, the associated Gaia sources occupy positions compatible with young to main sequence stars as well as stars of moderate age (in the few Gyr range). Differences between the SVM and Bayesian method 
are small and the identified Gaia sources are largely overlapping in the HRD. As expected, the distances towards the earlier spectral types tend to be larger than towards the later spectral types, in particular M dwarfs, which are found within about 200\,pc. Both methods associated stars on the red giant branch with eROSITA sources. These sources may not be causing the X-ray emission as red giants beyond a so-called dividing line have been found to lack genuine X-ray emission \citep{Haisch_1991}. We note, however, that (a) lower-mass companions may be responsible for the detected X-ray emission and (b) the association likelihoods of these sources are indeed quite low. In particular, the reddest giant in Fig.~\ref{fig:HRD} (HT Hya) is also a GALEX FUV and NUV source with $p_{stellar}\approx0.7$, i.e., this association may be very well be indeed
correct although the giant itself may not be the source of the X-ray emission. Nevertheless, we purposely keep these sources (and potentially other source classes) in the final sample to avoid biasing ourselves unduly towards current ``wisdom'' precluding new discoveries.

\begin{figure}[t]
  \includegraphics[width=0.498\textwidth]{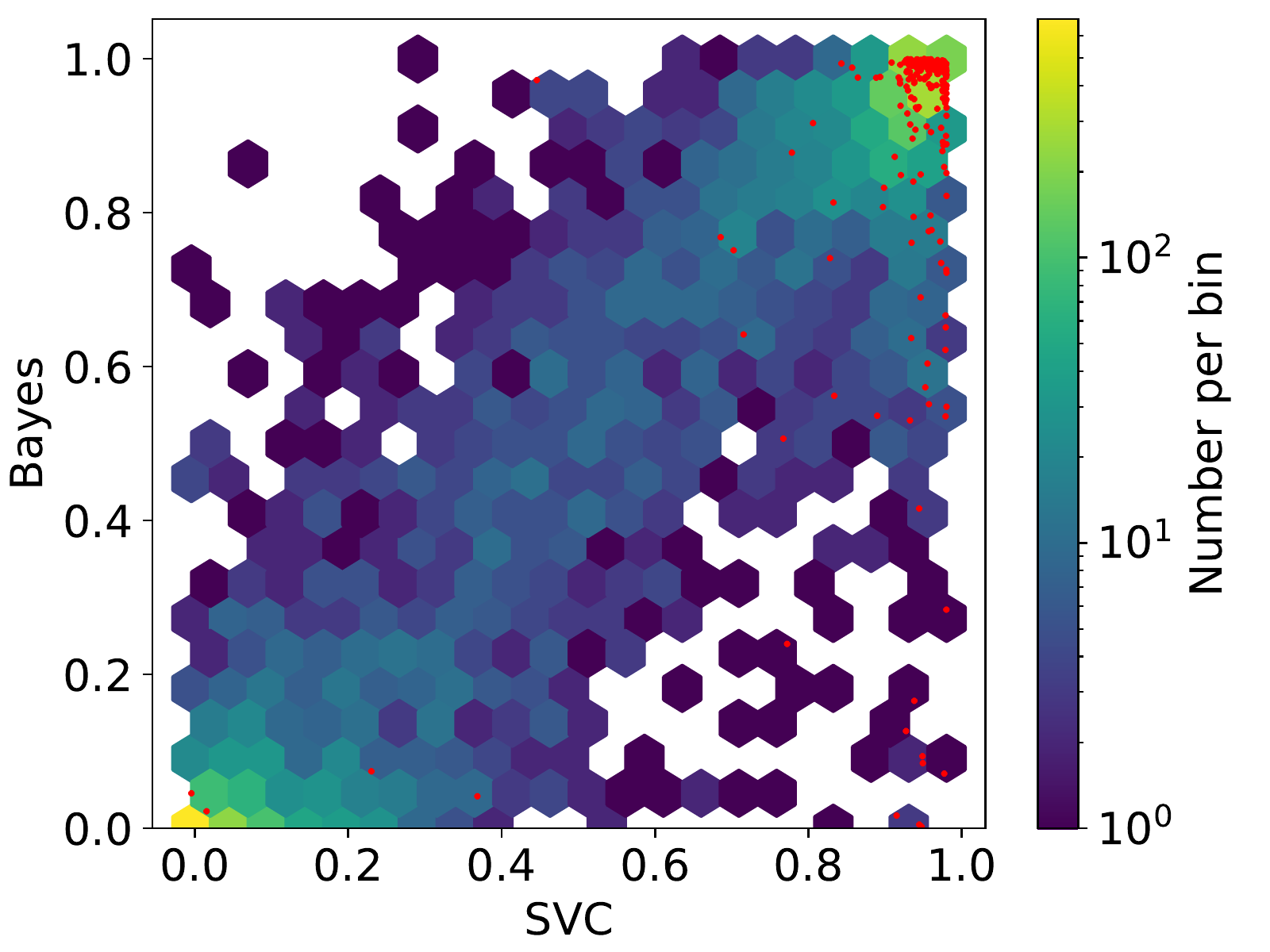}
  \caption{Association probabilities for the SVC and Bayesian methods. The red 
  dots depict the sources within 100\,pc.
  \label{fig:probs}}
\end{figure}

The parallax-distributions (Fig.~\ref{fig:log_plx}) show that the identified stellar sources peak between  log\,plx=0.5 and log\,plx=1, i.e., are mostly between 100\,pc and 250\,pc from the Sun. Few sources are farther than 1\,kpc (log\,plx<0). The objects associated by the Bayesian method and \citet{Salvato_2021} tend to have smaller parallaxes (larger distances) compared to the SVC. For the Bayesian method, this may be expected, because it is currently ignorant of the Gaia parallaxes. \citet{Salvato_2021} do consider the parallax in their the method, but it likely has a different (lower) importance than for the SVC, which can use the parallax information 
to boost the association  likelihood of certain associations (see red dots in Fig.~\ref{fig:probs}). In theory, the parallax information may be beneficially used to reduce 
the number of spurious associations as the number of possible random associations increases with distance. At the moment, however, the SVC is not 
able to take full advantage of this benefit resulting in similar reliability and completeness levels as the Bayesian approach, probably because of the limited training set.

Lastly, the match distance-distribution shows that the associations identified as stellar by all three 
methods have, on average, the smallest match distances (cf. Fig.~\ref{fig:match_dist}). 
The stellar associations shared between the Bayesian and SVC methods have relatively small match distances, too. 
The 
other panels of Fig.~\ref{fig:match_dist}, however,  show that the match distances tend to be
larger for associations found by only one method. For these associations, the positional match
is insufficient to result in a secure stellar identification and the weighting of the other 
features becomes important, which is different between the three methods.

\subsection{Probabilities}
Our Bayesian and SVC methods were developed focusing on stars (coronal emitters). Therefore,
we compare the calculated probabilities from the SVC and Bayesian methods in Fig.~\ref{fig:probs}
(after calibrating 
the SVC probabilities as described in sect.~\ref{sect:SVC_results}). Overall, 
the probabilities are very similar being clustered around the 1:1 relation, most sources get either very low 
or high association probabilities in both methods. In addition to some intrinsic scatter, 
a number of sources get relatively high association probabilities by the SVC method while the 
Bayesian method assigns only mediocre probabilities between 0.5 and 0.8 (the structure located in the right middle part 
of Fig.~\ref{fig:probs}. Most of these sources are within 100\,pc (red dots in Fig.~\ref{fig:probs}) so that 
they get boosted association probabilities due to these high parallax-values compared to 
the Bayesian approach, which currently ignores the parallaxes.

\subsection{HamStar Catalog}
The catalog of the stellar sources is based on the Bayesian framework, because it
directly provides probabilities reflecting our prior knowledge and the data. 
We also include two columns that indicate if the particular association is also 
identifified by the SVM approach and by NWAY \citep{Salvato_2021}. This catalog
is dubbed ``HamStars'' in the eROSITA context.

\section{Conclusions and Outlook \label{sect:conclusions}}
We present SVC and  Bayesian methods specifically designed to identify stars among the eROSITA sources. 
Both methods provide very similar results, both, in terms of sample quality (completeness and reliability) 
and resulting physical properties. 
Both methods provide significant improvements over a purely geometric approach reducing the number of spurious 
associations in the sample by about a factor of two and samples constructed to contain the geometrically expected number of 
eROSITA sources achieve almost 90\,\% completeness and reliability. Furthermore, we show how 
to construct calibrated
probabilities for both methods, which can be eventually used to create sub-samples with specific 
completeness and reliability properties by applying appropriate cuts in association probability.

On the one hand, it is somewhat surprising that the SVC and Bayesian methods perform so similarly, 
because the SVC method uses the parallax measurement as an additional information compared to the Bayesian
approach. On the other hand, the SVC method is ignorant of the ``mathematics''  of the matching distance and needs to 
learn the positional match characteristics from the training sample.
We expect that a larger training sample will, 
at least partly, mitigate this issue, in particular when applied to the eROSITA all-sky survey (eRASS).

The sample quality in terms of completeness and reliability is specific to eFEDS, because these properties
depend on (a) the depth of the X-ray exposure, (b) the 
ratio between stellar and other X-ray sources, and (c) the sky density of the eligible stellar counterparts. 
These properties are expected to change strongly with galactic position, e.g. the sky density of eligible stellar counterparts differs by at least a factor of hundred between the galactic pole and bulge regions. Therefore, these effects need to be taken into account to achieve similar or better results for eRASS compared to eFEDS. In addition, 
In the future, we plan to also include additional
information from the X-ray data such as spectral hardness to further improve the algorithms for their 
application to eRASS and it is straight forward to include additional information in both methods.
Lastly, future Gaia data releases will improve the quality of the match catalog. We are therefore confident that it is possible to construct well characterized stellar samples from eROSITA data. This large sample of stars with well known X-ray properties will allow us to improve our understanding of stellar activity throughout space and time.

\begin{acknowledgements} 
eROSITA  is  the  primary  instrument  aboard  SRG,  a  joint Russian-German  science  mission  supported  by  the  Russian  Space  Agency (Roskosmos), in the interests of the Russian Academy of Sciences represented by its Space Research Institute (IKI), and the Deutsches Zentrum für Luft- und Raumfahrt  (DLR).  The  SRG  spacecraft  was  built  by  Lavochkin  Association (NPOL)  and  its  subcontractors,  and  is  operated  by  NPOL  with  support  from IKI and the Max Planck Institute for Extraterrestrial Physics (MPE). The development and construction of the eROSITA X-ray instrument was led by MPE, with  contributions  from  the  Dr.  Karl  Remeis  Observatory  Bamberg  \&  ECAP (FAU  Erlangen-N\"urnberg),  the University  of  Hamburg  Observatory,  the Leibniz Institute for Astrophysics Potsdam (AIP), and the Institute for Astronomyand Astrophysics of the University of T\"ubingen, with the support of DLR and the Max Planck Society.  The Argelander Institute for Astronomy of the University of Bonn and the Ludwig Maximilians Universit\"at Munich also participated in the science preparation for eROSITA.

This work has made use of data from the European Space Agency (ESA) mission
{\it Gaia} (\url{https://www.cosmos.esa.int/gaia}), processed by the {\it Gaia}
Data Processing and Analysis Consortium (DPAC,
\url{https://www.cosmos.esa.int/web/gaia/dpac/consortium}). Funding for the DPAC
has been provided by national institutions, in particular the institutions
participating in the {\it Gaia} Multilateral Agreement.

PCS gratefully acknowledges support by the DLR under 50~OR~1901 and 50~OR~2102. 
SF acknowledge supports through the Integrationsamt Hildesheim, the ZAV of Bundesagentur f\"ur Arbeit, and the Hamburg University and thanks Gabriele Uth and Maria Theresa Lehmann for their support.
\end{acknowledgements}

\bibliographystyle{aa}
\bibliography{eroML}

\appendix
\section{Isochrones \label{sect:isochrones}}
\begin{figure}
  \centering
  \includegraphics[width=0.49\textwidth]{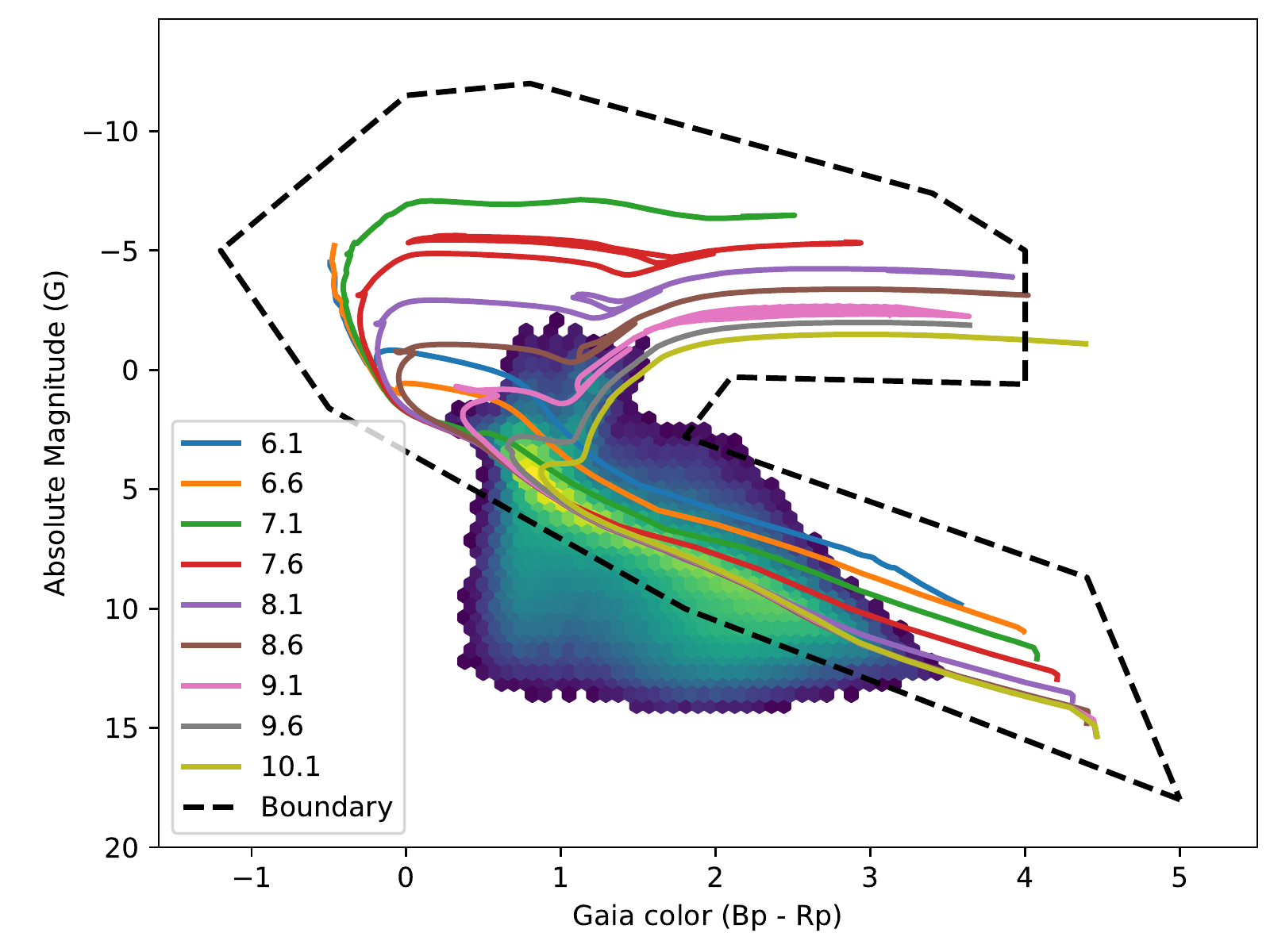}
  \caption{Polygon containing valid stellar sources with PARSEC evolutionary
    tracks. The 2D histogram in the background shows a typical Gaia source
    population; most sources are close to the 100+ Myrs main sequence.
  \label{fig:isochrones}}
\end{figure}

Figure~\ref{fig:isochrones} shows the region in which a Gaia source is 
considered an eligible stellar counterpart.

\end{document}